\documentclass[10pt,conference]{IEEEtran}
\IEEEoverridecommandlockouts

\usepackage{graphicx} 
\usepackage{booktabs}
\usepackage{multirow}
\usepackage{array}
\usepackage{geometry}
\usepackage{amsmath} 
\usepackage[caption=false,font=footnotesize]{subfig}
\usepackage{float}
\usepackage[ruled,vlined,linesnumbered]{algorithm2e}
\usepackage{comment}
\usepackage{hyperref}

\title{Lite-BD: A Lightweight Black-box Backdoor Defense via Reviving Multi-Stage Image Transformations}

\author{
\IEEEauthorblockN{Abdullah Arafat Miah\IEEEauthorrefmark{1},
Yu Bi\IEEEauthorrefmark{1}}
\IEEEauthorblockA{\IEEEauthorrefmark{1}
Department of Electrical, Computer, and Biomedical Engineering\\
University of Rhode Island, Kingston, RI, USA\\
\{abdullaharafat.miah, yu\_bi\}@uri.edu
}
}

\begin{document}

\maketitle

\begin{abstract}

Deep Neural Networks (DNNs) are vulnerable to backdoor attacks. Due to the nature of Machine Learning as a Service (MLaaS) applications, black-box defenses are more practical than white-box methods, yet existing purification techniques suffer from key limitations: a lack of justification for specific transformations, dataset dependency, high computational overhead, and a neglect of frequency-domain transformations. This paper conducts a preliminary study on various image transformations, identifying down-upscaling as the most effective backdoor trigger disruption technique. We subsequently propose \texttt{Lite-BD}, a lightweight two-stage blackbox backdoor defense. \texttt{Lite-BD} first employs a super-resolution-based down-upscaling stage to neutralize spatial triggers. A secondary stage utilizes query-based band-by-band frequency filtering to remove triggers hidden in specific bands. Extensive experiments against state-of-the-art attacks demonstrate that \texttt{Lite-BD} provides robust and efficient protection. Codes can be found at \url{https://github.com/SiSL-URI/Lite-BD}.
    
\end{abstract}


\section{Introduction}


 With the rise of Machine Learning as a Service (MLaaS) \cite{gu2019badnets}, Artificial Intelligence (AI) usage has become increasingly dependent on third parties, creating serious security vulnerabilities for end users, such as backdoor attacks \cite{li2021backdoor}. In a conventional backdoor attack scenario, an adversary inserts a backdoor into a deep model by poisoning it with a malicious dataset, either during training from scratch or during fine-tuning at any stage of the supply chain. Consequently, the model performs normally when clean samples are fed into it but will predict an adversarial target when trigger patterns appear in the inputs. To defend against backdoor attacks, several defense methods have been proposed; however, most require white-box access to the model \cite{li2021anti} \cite{wu2021adversarial}. While these methods can produce effective mitigation, their utility is limited for MLaaS systems where the end user lacks access to model parameters. To address this gap, researchers have proposed black-box backdoor defenses—such as backdoor sample detection \cite{liu2023detecting}, model detection \cite{xu2024ban}, and sample purification \cite{Shi2023ZIP, Yang2024SampDetox}, where the defender has no access to the model internals. Despite the promises of sample purification based methods, they face several limitations: \textit{i) there is often no proper justification for the choice of specific spatial transformations, ii) heavily adopted diffusion models may require costly task-specific training introducing significant computational overhead, and iii) if spatial transformations fail to disrupt a specific trigger, the potential for frequency filtering as an alternative remains largely unexplored.}


To this end, we first evaluate various spatial transformations through a preliminary study, identifying downscaling-then-upscaling as the most effective trigger disruption method. We then propose \textbf{\texttt{Lite-BD}}, a novel and lightweight black-box trigger purification method. In the upscaling phase, we employ neural super-resolution \cite{wang2021realesrgan, liang2021swinir} to provide effective image restoration without the high computational cost of diffusion models. Our zero-shot approach eliminates dataset dependency and training overhead. For triggers that bypass spatial transformations, we further propose a band-by-band frequency filtering technique to neutralize signatures in the frequency domain. In summary, this paper makes the following three main contributions:

\begin{itemize}
    \item We evaluate various spatial and inverse transformations to identify optimal configurations for disrupting backdoor triggers.
    \item We propose \textbf{\texttt{Lite-BD}}, a novel two-stage black-box backdoor defense framework. The first stage purifies poisoned samples through a pipeline of downscaling and neural super-resolution-based upscaling; if the trigger persists, a second stage employs query-based band-by-band frequency filtering to neutralize triggers in the frequency domain.
    \item Through extensive experiments, we demonstrate that \textbf{\texttt{Lite-BD}} effectively reduces Attack Success Rate (ASR) with minimal impact on benign performance, outperforming existing black-box methods in computational efficiency.
\end{itemize}

\begin{figure}
    \centering
    \includegraphics[width =\linewidth]{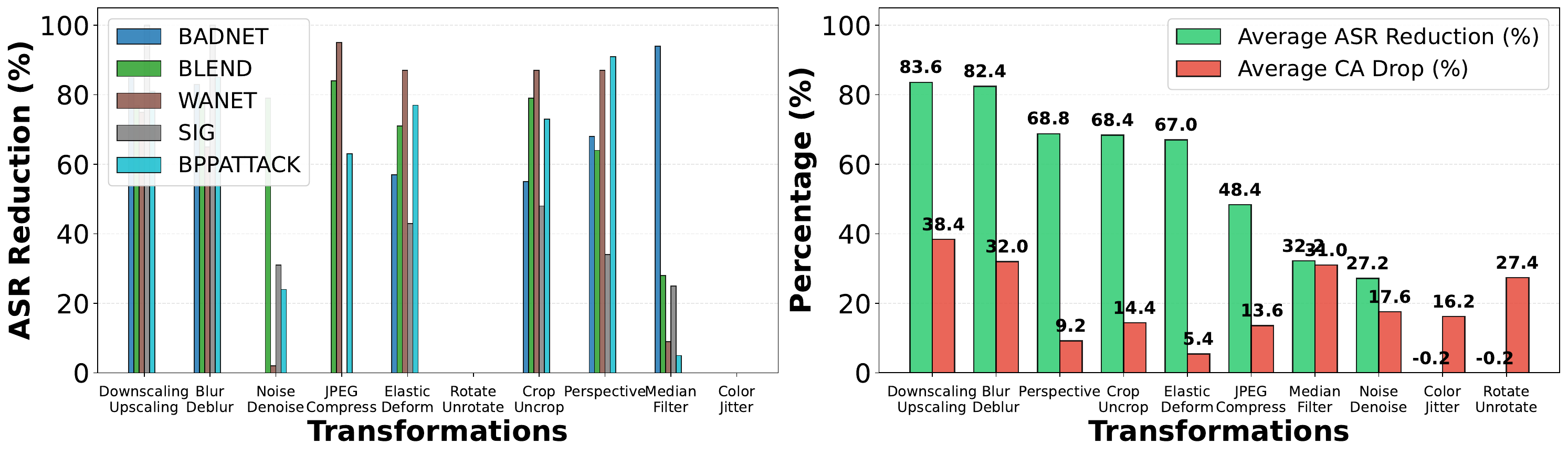}
    \label{fig:preli_study}
    \caption{Spatial transformation effects on backdoor attacks. The left panel shows ASR reduction across all five attacks and transformations. The right panel presents average performance metrics sorted by ASR reduction.}
\end{figure}

\vspace{-5mm}

\section{Related Works}

\vspace{-1pt}

\noindent \textbf{Backdoor Attacks.} Existing backdoor attacks can be broadly categorized into those using visible and invisible triggers. For example, BadNets \cite{gu2019badnets} employs patch-based visible trigger patterns, whereas Blend \cite{chen2017targeted} and WaNet \cite{nguyen2021wanet} utilize invisible triggers. Triggers can also be adaptive, such as LIRA \cite{doan2021lira}, or image-quantization-based, as in BppAttack \cite{wang2022bppattack}. Backdoor attacks may operate in the frequency domain, such as FIBA \cite{feng2022fiba} and LF \cite{li2020rethinking}, or jointly in both the spatial and frequency domains, as demonstrated by DUBA \cite{gao2024dual}. More recently, attacks have been extended to simultaneously exploit the spatial, frequency, and feature domains, as in \cite{gao2025triple}. Additionally, some attacks can be mounted without label poisoning, including SIG \cite{barni2019new} and LC \cite{turner2018clean}.

\noindent \textbf{Backdoor Defense.} In white-box backdoor defense, the defender has access to the model and can control the training process \cite{li2021anti}, perform fine-tuning \cite{zhao2024defending,zhu2023enhancing}, or apply pruning-based strategies \cite{liu2018fine,wu2021adversarial,li2023reconstructive}. In the black-box setting, backdoors can be mitigated through backdoored model detection \cite{xiang2023cbd}, poisoned sample detection \cite{liu2023detecting}, or sample purification methods that aim to destroy triggers at test time, such as Sancdifi \cite{may2023salient}, BDMAE \cite{sun2023mask}, ZIP \cite{Shi2023ZIP}, and SampDetox \cite{Yang2024SampDetox}. However, existing purification-based black-box defenses suffer from several key limitations, including effectiveness restricted to local triggers, strong dataset dependency, and additional computational overhead.

\begin{figure*} [h]
    \centering
    \includegraphics[width=\linewidth]{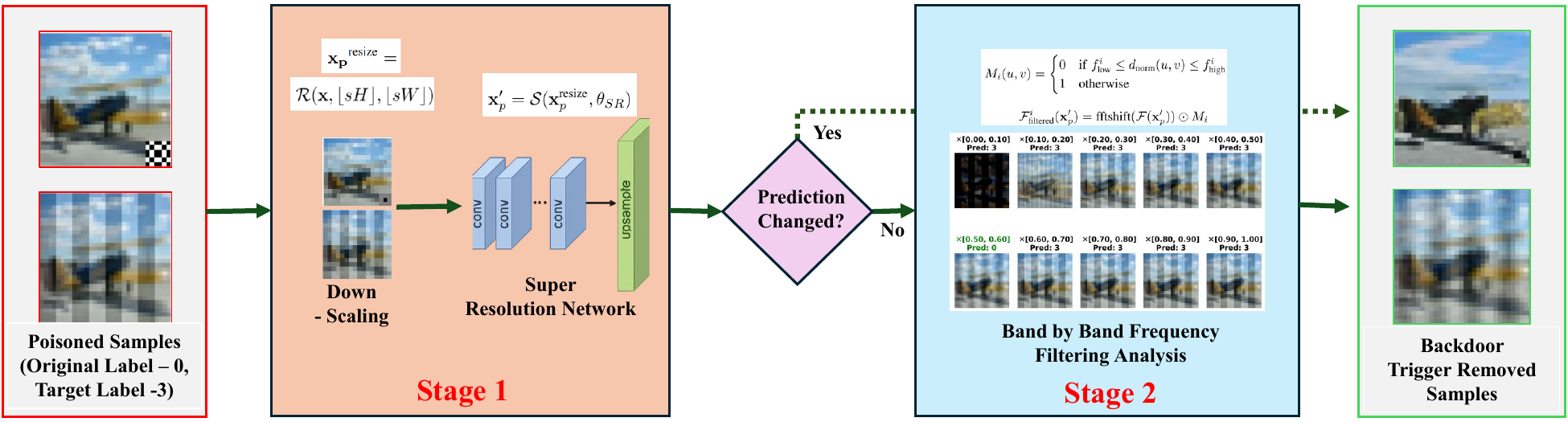}
    \caption{Overview of the Proposed Black-box Backdoor Defense \texttt{Lite-BD}.}
    \label{fig:overview}
\end{figure*}

\vspace{-3mm}

\section{Preliminary Study}

To evaluate the efficacy of spatial transformations in disrupting backdoor triggers, we conduct an initial study using five representative attacks: BadNets \cite{gu2019badnets}, Blend \cite{chen2017targeted}, WaNet \cite{nguyen2021wanet}, SIG \cite{barni2019new}, and BppAttack \cite{wang2022bppattack}. Using ResNet-18 on CIFAR-10, we measure Attack Success Rate (ASR) and Clean Accuracy (CA) across 100 random samples and their poisoned counterparts before and after ten distinct transformation-inverse pairs. Evaluated operations include: \textit{Down-Upscaling} (50\% bilinear), \textit{Blur-Deblur} ($5 \times 5$ Gaussian kernel/unsharp mask), \textit{Noise-Denoise} (Gaussian noise/Non-Local Means), \textit{JPEG Compression} (quality 50), \textit{Elastic Deformation}, \textit{Rotate-Unrotate} ($15^{\circ}$), \textit{Crop-Uncrop}, \textit{Perspective Transformation}, \textit{Median Filtering}, and \textit{Color Jittering}. Results in Figure~\ref{fig:preli_study} indicate that \textit{Downscaling-Upscaling} achieves the highest average ASR reduction, followed by \textit{Blur-Deblur}. While ShrinkPad~\cite{li2021backdoor} utilizes image shrinking, its reliance on padding significantly degrades CA. Recent defenses like ZIP~\cite{Shi2023ZIP} and SampDetox~\cite{Yang2024SampDetox} employ \textit{Blur-Deblur} and \textit{Noise-Denoise} respectively, but rely on computationally expensive diffusion models for restoration.

\vspace{-5pt}

\section{Methodology}

\subsection{Threat Model}

We assume a black-box defense scenario where the defender lacks access to model parameters and/or fine-tuning capabilities, and is restricted to observing output predictions. The defender's objective is to filter poisoned inputs and purify samples to disrupt triggers while maintaining benign semantics. Conversely, we assume a strong adversary with full control over the training process, capable of injecting triggers that cause malicious misclassification while preserving normal performance on clean inputs.

\vspace{-3mm}

\subsection{Overview}

Based on the observations from the preliminary study, we propose a two-stage black-box defense framework \texttt{Lite-BD} designed to disrupt backdoor triggers in poisoned samples and purify them for correct classification into their benign labels. In the first stage, we implement a "Downscaling-Upscaling" operation by employing stochastic resizing followed by neural super-resolution. If this operation fails to disrupt the trigger, a condition determined by auxiliary queries that fails to show a label flip, the output is passed to the second stage. The subsequent stage utilizes a band-by-band frequency filtering technique specifically designed to destroy triggers that remain robust against the initial resizing operations. If neither stage results in a label flip, the input image is deemed backdoor-free. An overview of our proposed \texttt{Lite-BD} is given in Figure \ref{fig:overview}.

\subsection{Stage 1: Stochastic Resize and Neural Super-Resolution}

For a given poisoned image $\mathbf{x}_p \in \mathbf{R}^{H \times W \times C}$, where $H$, $W$, and $C$ represent the image dimensions, we apply a stochastic downscaling operation with a scaling factor $s \in [s_{\min}, s_{\max}]$, where $s_{\min}, s_{\max} \in (0, 1)$. The dimensions of the downscaled image are computed via Equation \ref{eq:downscale}, where $\mathcal{B}(\cdot)$ denotes bilinear interpolation. Such geometric transformation effectively disrupts most backdoor triggers, as trigger patterns often rely on precise pixel-level placement.

\vspace{-5mm}

\begin{equation} \label{eq:downscale}
    \mathbf{x_p}^{\text{resize}} = \mathcal{B}(\mathbf{x}, \lfloor sH \rfloor, \lfloor sW \rfloor)
\end{equation}

For a benign sample $\mathbf{x}_c$ and its poisoned counterpart $\mathbf{x}_p$, the downscaling operation yields $\mathbf{x}_p^{\text{resize}}$. Our objective is to apply an inverse transformation to restore the image to its original dimensions $H \times W$. Let the reconstructed image be denoted as $\mathbf{x}'_p$. For ideal purification, the condition $\mathbf{x}'_p = \mathbf{x}_c$ should be satisfied. To achieve this, we employ a pretrained super-resolution (SR) network, denoted as $\mathcal{S}(\cdot)$, to perform the upscaling:

\vspace{-8pt}

\begin{equation} \label{eq:upscale}
    \mathbf{x}'_p = \mathcal{S}(\mathbf{x}_p^{\text{resize}}, \theta_{SR})
\end{equation}

\noindent where $\theta_{SR}$ represents the parameters of the pretrained backbone. Specifically, we utilize the Enhanced Super-Resolution Generative Adversarial Network (Real-ESRGAN \cite{wang2021realesrgan}) with $4\times$ upscaling and the Swin Transformer-based architecture (SwinIR \cite{liang2021swinir}) with $2\times$ upscaling which are referred to as \textbf{\texttt{Lite-BD (RE)}} and \textbf{\texttt{Lite-BD (SW)}}, respectively. During downscaling, aggregating source pixels disrupts the spatial coherence of pixel-level triggers. Pretrained super-resolution networks then reconstruct semantic information using learned natural image priors rather than local artifacts. Because trigger patterns are typically out-of-distribution (OOD) for SR networks, they are not recovered during upscaling. Consequently, the purified image $\mathbf{x}'_p$ retains semantic integrity while the malicious trigger is neutralized.

\subsection{Stage 2: Band-by-Band Frequency Filtering}

Following Stage 1, the purification of a potentially poisoned sample is verified by querying the target model. If the backdoor trigger has been successfully disrupted, the predicted label should no longer be flipped to the target class. However, if the trigger remains intact, the sample will proceed to Stage 2, where we conduct a band-by-band frequency filtering analysis. For a given input image from Stage 1, $\mathbf{x}'_p \in \mathbf{R}^{C \times H \times W}$, we apply a 2D Discrete Fourier Transform (DFT) to each color channel via Equation \ref{eq:fourier}, where $(u, v)$ represents the frequency domain coordinates. The resulting coefficients are shifted to center the zero-frequency component (DC component). We then perform frequency normalization using Equation \ref{eq:freq_normalization} to map distances from the center:

\vspace{-5mm}

\begin{equation}  \label{eq:fourier}
\mathcal{F}(\mathbf{x}'_p)(u,v) = \sum_{h=0}^{H-1} \sum_{w=0}^{W-1} \mathbf{x}'_p(h,w) \cdot e^{-j2\pi\left(\frac{uh}{H} + \frac{vw}{W}\right)}
\end{equation}

\vspace{-6mm}

\begin{equation} \label{eq:freq_normalization}
d_{\text{norm}}(u, v) = \frac{\sqrt{(u - u_c)^2 + (v - v_c)^2}}{\sqrt{u_c^2 + v_c^2}}
\end{equation}

The normalized frequency spectrum is partitioned into $n$ uniformly spaced concentric bands, where each band $B_i$ is defined by the range $[f_{\text{low}}^i, f_{\text{high}}^i]$. For each band, we apply a band-stop filter mask $M_i$ as defined in Equation \ref{eq:filter_mask}. The filtered frequency representation is obtained through element-wise multiplication (Hadamard product) in Equation \ref{eq:band_filter}, and the filtered image is reconstructed using the Inverse Fourier Transform in Equation \ref{eq:freq_reverse}, where $\mathcal{R}\{\cdot\}$ extracts the real component:

\vspace{-8pt}

\begin{equation} \label{eq:filter_mask}
M_i(u, v) = \begin{cases} 
0 & \text{if } f_{\text{low}}^i \leq d_{\text{norm}}(u, v) \leq f_{\text{high}}^i \\
1 & \text{otherwise}
\end{cases}
\end{equation}

\vspace{-4mm}

\begin{equation} \label{eq:band_filter}
\mathcal{F}_{\text{filtered}}^i(\mathbf{x}'_p) = \text{fftshift}(\mathcal{F}(\mathbf{x}'_p)) \odot M_i
\end{equation}

\vspace{-6mm}

\begin{equation} \label{eq:freq_reverse}
\tilde{\mathbf{x}}_p^{\prime i} = \mathcal{R}\{\mathcal{F}^{-1}(\text{ifftshift}(\mathcal{F}_{\text{filtered}}^i(\mathbf{x}'_p)))\}
\end{equation}


This process generates $n$ filtered candidates $\{\tilde{\mathbf{x}}_p^{\prime 1}, \dots, \tilde{\mathbf{x}}_p^{\prime n}\}$ from a single image $\mathbf{x}'_p$. We pass each candidate through the model to detect restored labels, indicating the trigger's frequency band. If multiple bands disrupt the trigger, we prioritize higher frequencies to preserve semantic integrity. The optimal candidate then undergoes bilateral filtering or unsharp masking to restore edge sharpness. If both stages fail to correct the label flip, the sample is considered clean, and the original image is returned. 

\begin{figure*}
    \centering
    \includegraphics[width=0.9\linewidth]{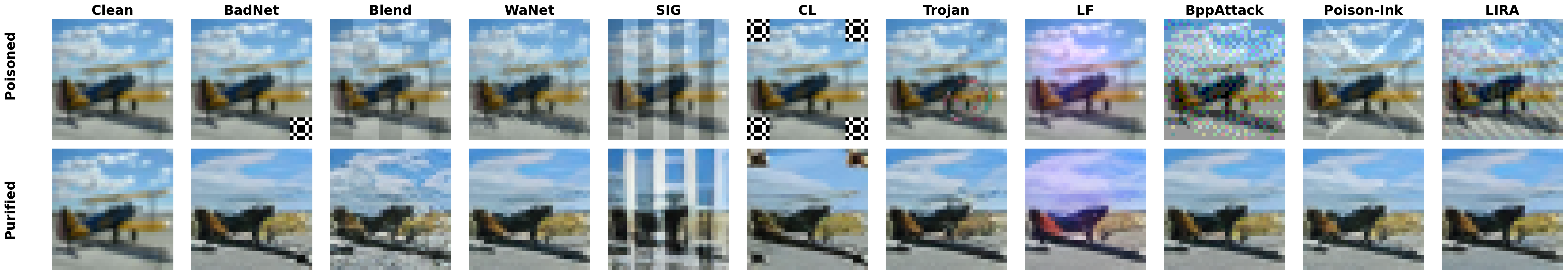}
    \caption{Illustration of poisoned samples and their corresponding purified samples across ten backdoor attacks using our proposed method with \texttt{Lite-BD (RE)}.}
    \label{fig:purified_samples_visulization}
\end{figure*}

\begin{table*}[htbp]
\centering
\tiny
\caption{Defense Performance Comparison. \textbf{Bold} denotes the best result; \underline{underline} denotes the second-best.}
\label{tab:defense_comparison}
\resizebox{\textwidth}{!}{%
\begin{tabular}{@{}llccccccccccccccccccccc@{}}
\toprule
\multirow{2}{*}{Dataset} & \multirow{2}{*}{Attack} & \multirow{2}{*}{Base CA} & \multirow{2}{*}{Base ASR} & \multicolumn{3}{c}{ShrinkPad} & \multicolumn{3}{c}{ZIP} & \multicolumn{3}{c}{Sampdetox} & \multicolumn{3}{c}{\texttt{Lite-BD (RE)}} & \multicolumn{3}{c}{\texttt{Lite-BD (SW)}} \\
\cmidrule(lr){5-7} \cmidrule(lr){8-10} \cmidrule(lr){11-13} \cmidrule(lr){14-16} \cmidrule(lr){17-19}
& & & & CA & PA & ASR & CA & PA & ASR & CA & PA & ASR & CA & PA & ASR & CA & PA & ASR \\
\midrule
\multirow{11}{*}{\rotatebox{90}{CIFAR-10}} 
& BadNet & 0.950 & 1.000 & 0.699 & 0.386 & 0.495 & \textbf{0.886} & 0.022 & 1.000 & 0.763 & 0.034 & 0.988 & \underline{0.865} & \textbf{0.911} & \underline{0.024} & 0.847 & \underline{0.897} & \textbf{0.024} \\
& Blend & 0.946 & 0.999 & 0.731 & 0.350 & 0.299 & \underline{0.881} & 0.724 & 0.064 & 0.772 & 0.648 & \underline{0.063} & \textbf{0.885} & \textbf{0.838} & \textbf{0.023} & 0.861 & \underline{0.659} & 0.164 \\
& WaNet & 0.945 & 1.000 & 0.556 & 0.395 & 0.592 & \underline{0.879} & 0.827 & \underline{0.054} & 0.788 & 0.737 & \underline{0.054} & \textbf{0.886} & \textbf{0.877} & 0.052 & 0.870 & \underline{0.857} & \textbf{0.041} \\
& SIG & 0.866 & 0.999 & 0.742 & 0.361 & \textbf{0.001} & 0.878 & 0.079 & 0.920 & 0.779 & 0.309 & \underline{0.307} & \textbf{0.888} & \underline{0.428} & 0.356 & \underline{0.867} & \textbf{0.516} & \underline{0.011} \\
& CL & 0.865 & 1.000 & 0.732 & 0.364 & \underline{0.025} & 0.871 & 0.022 & 1.000 & 0.794 & 0.084 & 0.938 & \textbf{0.889} & \textbf{0.879} & \textbf{0.000} & \underline{0.875} & \underline{0.866} & \textbf{0.000} \\
& BPPAttack & 0.945 & 1.000 & 0.727 & 0.345 & 0.588 & \underline{0.878} & \underline{0.832} & \underline{0.047} & 0.774 & 0.769 & \underline{0.044} & \textbf{0.881} & \textbf{0.903} & \textbf{0.013} & 0.863 & 0.767 & 0.058 \\
& Trojan & 0.947 & 1.000 & 0.684 & 0.470 & 0.230 & \underline{0.877} & \underline{0.786} & \textbf{0.047} & 0.790 & 0.702 & \underline{0.054} & \textbf{0.881} & \textbf{0.806} & \underline{0.072} & 0.858 & 0.710 & 0.084 \\
& LF & 0.944 & 0.997 & 0.675 & 0.122 & 0.968 & \underline{0.880} & 0.254 & 0.748 & 0.781 & \underline{0.511} & \underline{0.400} & \textbf{0.884} & \textbf{0.893} & \textbf{0.016} & 0.870 & \textbf{0.893} & \textbf{0.016} \\
& Poison-Ink & 0.946 & 0.998 & 0.677 & 0.294 & 0.535 & \textbf{0.880} & \textbf{0.828} & \underline{0.057} & 0.769 & \underline{0.753} & \textbf{0.046} & \underline{0.873} & \underline{0.779} & 0.148 & 0.864 & 0.631 & 0.209 \\
& LIRA & 0.941 & 1.000 & 0.738 & 0.517 & 0.170 & \underline{0.875} & \underline{0.746} & \underline{0.102} & 0.783 & 0.736 & \underline{0.042} & \textbf{0.887} & \textbf{0.890} & \textbf{0.014} & 0.852 & 0.567 & 0.224 \\
\cmidrule(l){2-19}
& \textbf{Average} & 0.933 & 0.999 & 0.696 & 0.360 & 0.390 & \underline{0.879} & 0.512 & \underline{0.404} & 0.779 & 0.498 & \underline{0.294} & \textbf{0.882} & \textbf{0.821} & \textbf{0.072} & 0.863 & \underline{0.736} & 0.083 \\
\midrule
\multirow{11}{*}{\rotatebox{90}{Fashion-MNIST}} 
& BadNet & 0.952 & 1.000 & 0.279 & 0.351 & 0.497 & \textbf{0.897} & 0.000 & 1.000 & 0.256 & 0.256 & 0.707 & \underline{0.776} & \textbf{0.807} & \textbf{0.038} & \textbf{0.845} & \textbf{0.807} & \underline{0.113} \\
& Blend & 0.950 & 0.999 & 0.268 & 0.266 & \underline{0.164} & \textbf{0.857} & \underline{0.684} & 0.191 & \underline{0.837} & \textbf{0.773} & \textbf{0.055} & 0.754 & 0.616 & \underline{0.076} & 0.803 & 0.539 & 0.133 \\
& WaNet & 0.949 & 1.000 & 0.175 & 0.109 & 0.989 & \textbf{0.893} & \textbf{0.835} & \textbf{0.027} & \underline{0.866} & \underline{0.854} & \textbf{0.022} & 0.691 & 0.602 & 0.273 & 0.757 & 0.818 & \underline{0.034} \\
& SIG & 0.860 & 0.999 & 0.453 & 0.172 & \textbf{0.000} & \textbf{0.857} & 0.086 & 0.892 & \underline{0.828} & \textbf{0.624} & \underline{0.278} & 0.770 & \underline{0.560} & \underline{0.095} & 0.820 & 0.551 & 0.105 \\
& CL & 0.858 & 1.000 & 0.397 & 0.151 & \textbf{0.000} & \textbf{0.933} & 0.001 & 0.999 & \underline{0.909} & \underline{0.192} & \underline{0.803} & 0.783 & \textbf{0.679} & \textbf{0.197} & 0.855 & \underline{0.370} & \underline{0.620} \\
& BPPAttack & 0.949 & 0.998 & 0.070 & 0.100 & 1.000 & \textbf{0.926} & \underline{0.869} & 0.060 & \underline{0.912} & \textbf{0.908} & \textbf{0.017} & 0.547 & 0.795 & \textbf{0.017} & 0.592 & 0.870 & \textbf{0.011} \\
& Trojan & 0.950 & 1.000 & 0.285 & 0.312 & 0.427 & \textbf{0.867} & \textbf{0.848} & \underline{0.014} & \underline{0.854} & \underline{0.830} & \underline{0.017} & 0.758 & 0.781 & \textbf{0.006} & 0.835 & 0.527 & 0.070 \\
& LF & 0.950 & 1.000 & 0.297 & 0.165 & 0.849 & \underline{0.667} & \underline{0.632} & \underline{0.318} & 0.605 & 0.585 & 0.358 & 0.765 & \textbf{0.713} & \textbf{0.000} & \textbf{0.822} & \underline{0.712} & \underline{0.009} \\
& Poison-Ink & 0.952 & 1.000 & 0.272 & 0.148 & 0.414 & \textbf{0.902} & \textbf{0.907} & \underline{0.021} & \underline{0.889} & \underline{0.892} & \textbf{0.019} & 0.772 & 0.427 & 0.415 & 0.842 & 0.449 & 0.442 \\
& LIRA & 0.952 & 0.991 & 0.312 & 0.249 & 0.171 & \textbf{0.871} & \underline{0.817} & \underline{0.071} & 0.850 & \underline{0.817} & \underline{0.058} & 0.722 & 0.435 & \textbf{0.029} & \underline{0.761} & 0.397 & 0.212 \\
\cmidrule(l){2-19}
& \textbf{Average} & 0.937 & 0.999 & 0.281 & 0.202 & 0.451 & \textbf{0.867} & \underline{0.568} & \underline{0.369} & 0.843 & \textbf{0.713} & \underline{0.233} & 0.734 & \underline{0.642} & \textbf{0.113} & \underline{0.793} & 0.599 & \underline{0.175} \\
\midrule
\multirow{11}{*}{\rotatebox{90}{GTSRB}} 
& BadNet & 0.947 & 0.937 & 0.091 & 0.064 & \underline{0.162} & \underline{0.942} & \textbf{0.604} & \underline{0.506} & 0.892 & \underline{0.809} & \underline{0.242} & 0.912 & 0.436 & 0.544 & \textbf{0.951} & 0.519 & \textbf{0.465} \\
& Blend & 0.952 & 0.997 & 0.055 & 0.055 & \textbf{0.007} & 0.947 & 0.360 & 0.774 & 0.892 & \textbf{0.797} & \underline{0.237} & \underline{0.958} & \underline{0.657} & \textbf{0.177} & \textbf{0.977} & \underline{0.613} & \underline{0.221} \\
& WaNet & 0.949 & 0.999 & 0.060 & 0.042 & 0.398 & \textbf{0.947} & \textbf{0.928} & \underline{0.141} & 0.891 & 0.887 & \underline{0.132} & 0.613 & 0.716 & 0.218 & \underline{0.744} & \underline{0.926} & \textbf{0.038} \\
& SIG & 0.896 & 0.680 & 0.086 & 0.038 & \textbf{0.000} & \underline{0.832} & 0.413 & \underline{0.697} & 0.782 & \underline{0.451} & 0.622 & 0.777 & \underline{0.615} & \textbf{0.026} & \textbf{0.847} & \textbf{0.626} & \underline{0.040} \\
& CL & 0.910 & 0.480 & 0.083 & 0.071 & \underline{0.010} & \textbf{0.935} & \textbf{0.921} & \underline{0.130} & 0.886 & 0.880 & \underline{0.129} & 0.855 & \underline{0.882} & \underline{0.007} & \underline{0.917} & \underline{0.916} & \textbf{0.004} \\
& BPPAttack & 0.921 & 0.933 & 0.089 & 0.056 & \textbf{0.036} & \textbf{0.937} & \textbf{0.931} & \underline{0.125} & 0.873 & 0.889 & \underline{0.138} & 0.751 & 0.853 & \underline{0.044} & \underline{0.796} & \underline{0.857} & \underline{0.093} \\
& Trojan & 0.931 & 0.998 & 0.114 & 0.050 & 0.304 & \textbf{0.936} & \textbf{0.875} & \underline{0.165} & \underline{0.878} & \underline{0.844} & \textbf{0.123} & 0.687 & 0.527 & \underline{0.125} & \underline{0.917} & 0.096 & 0.830 \\
& LF & 0.927 & 0.994 & 0.069 & 0.051 & \underline{0.066} & \textbf{0.929} & 0.239 & 0.876 & 0.878 & \underline{0.527} & \underline{0.510} & \underline{0.850} & \textbf{0.777} & \textbf{0.000} & 0.880 & \underline{0.775} & \textbf{0.000} \\
& Poison-Ink & 0.946 & 0.944 & 0.050 & 0.045 & \underline{0.221} & \underline{0.943} & \textbf{0.924} & \textbf{0.142} & 0.890 & 0.882 & \underline{0.132} & 0.934 & 0.627 & 0.339 & \textbf{0.974} & \underline{0.453} & 0.524 \\
& LIRA & 0.948 & 0.997 & 0.122 & 0.084 & 0.524 & \textbf{0.937} & \underline{0.830} & \underline{0.246} & 0.888 & \textbf{0.857} & \textbf{0.152} & \underline{0.922} & \underline{0.804} & \underline{0.515} & 0.957 & 0.447 & 0.555 \\
\cmidrule(l){2-19}
& \textbf{Average} & 0.933 & 0.876 & 0.082 & 0.056 & \textbf{0.173} & \textbf{0.929} & \underline{0.703} & 0.380 & 0.875 & \textbf{0.783} & \underline{0.242} & 0.826 & \underline{0.680} & \textbf{0.200} & \underline{0.906} & 0.623 & \underline{0.277} \\
\bottomrule
\end{tabular}%
}
\end{table*}

\begin{table*}[t]
\centering
\caption{Comparison of Execution Times (Seconds) for Each Sample. \textbf{Bold} denotes the fastest execution times.}
\label{tab:execution_time}
\resizebox{\textwidth}{!}{
\begin{tabular}{llccccccccccc}
\hline
\textbf{Dataset} & \textbf{Method} & \textbf{BadNet} & \textbf{Blend} & \textbf{WaNet} & \textbf{SIG} & \textbf{CL} & \textbf{Trojan} & \textbf{LF} & \textbf{Poison-Ink} & \textbf{BppAttack} & \textbf{LIRA} & \textbf{Average} \\ \hline
\multirow{4}{*}{CIFAR-10} & \texttt{Lite-BD (RE)} & 0.02 & \textbf{0.02} & 0.02 & 0.11 & 0.02 & 0.03 & \textbf{0.12} & \textbf{0.04} & \textbf{0.06} & \textbf{0.13} & \textbf{0.05} \\
  & \texttt{Lite-BD (SW)} & \textbf{0.02} & 0.05 & \textbf{0.02} & \textbf{0.04} & \textbf{0.02} & \textbf{0.03} & 0.12 & 0.04 & 0.06 & 0.18 & 0.06 \\
  & SampDetox & 5.12 & 5.15 & 4.79 & 4.96 & 5.02 & 5.55 & 4.91 & 4.83 & 9.42 & 11.96 & 6.17 \\
  & ZIP & 1.00 & 1.00 & 1.00 & 1.02 & 1.03 & 1.15 & 0.99 & 0.99 & 2.42 & 6.38 & 1.70 \\ \hline
\multirow{4}{*}{Fashion-MNIST} & \texttt{Lite-BD (RE)} & \textbf{0.02} & \textbf{0.08} & 0.04 & \textbf{0.12} & \textbf{0.02} & \textbf{0.03} & \textbf{0.12} & \textbf{0.12} & \textbf{0.05} & \textbf{0.14} & \textbf{0.07} \\
  & \texttt{Lite-BD (SW)} & 0.03 & 0.13 & \textbf{0.02} & 0.12 & 0.09 & 0.13 & 0.12 & 0.12 & 0.05 & 0.23 & 0.11 \\
  & SampDetox & 6.34 & 5.89 & 6.37 & 6.09 & 5.51 & 6.91 & 5.60 & 5.85 & 9.59 & 13.11 & 7.13 \\
  & ZIP & 1.08 & 1.02 & 1.07 & 1.09 & 1.05 & 1.09 & 0.93 & 1.02 & 2.02 & 3.18 & 1.35 \\ \hline
\multirow{4}{*}{GTSRB} & \texttt{Lite-BD (RE)} & 0.07 & \textbf{0.08} & 0.02 & 0.09 & 0.08 & \textbf{0.05} & 0.10 & \textbf{0.04} & \textbf{0.14} & 0.18 & 0.09 \\
  & \texttt{Lite-BD (SW)} & \textbf{0.07} & 0.10 & \textbf{0.02} & \textbf{0.03} & \textbf{0.08} & 0.06 & \textbf{0.10} & 0.07 & 0.15 & \textbf{0.18} & \textbf{0.08} \\
  & SampDetox & 4.25 & 4.61 & 4.42 & 4.41 & 4.28 & 4.95 & 4.15 & 4.01 & 15.02 & 8.97 & 5.91 \\
  & ZIP & 1.01 & 1.04 & 1.05 & 0.97 & 1.03 & 1.12 & 0.95 & 0.95 & 4.94 & 4.37 & 1.74 \\ \hline
\end{tabular}
}
\end{table*}

\begin{table*}[htbp]
\centering
\caption{Performance Comparison Considering Different Stages. \textbf{S1}: Stage 1, \textbf{S2}: Stage 2. This experiment is done by selecting random 1000 samples for each attack.}
\label{tab:comprehensive_ablation}
\resizebox{\textwidth}{!}{
\begin{tabular}{llcccccccc}
\toprule
\textbf{Dataset} & \textbf{Attack} & \textbf{Baseline CA} & \textbf{Baseline ASR} &
\textbf{S1 PA} & \textbf{S1 ASR} &
\textbf{S2 PA} & \textbf{S2 ASR} &
\textbf{S1 + S2 PA} & \textbf{S1 + S2 ASR} \\
\midrule

\multirow{11}{*}{CIFAR-10}
& BadNet     & 0.950 & 1.000 & 0.919 & 0.040 & 0.040 & 0.982 & 0.921 & 0.017 \\
& Blend      & 0.946 & 0.999 & 0.841 & 0.038 & 0.142 & 0.842 & 0.845 & 0.014 \\
& WaNet      & 0.945 & 1.000 & 0.878 & 0.066 & 0.039 & 0.982 & 0.893 & 0.043 \\
& SIG        & 0.866 & 0.999 & 0.044 & 0.953 & 0.632 & 0.032 & 0.484 & 0.226 \\
& CL         & 0.865 & 1.000 & 0.868 & 0.000 & 0.069 & 0.951 & 0.868 & 0.000 \\
& BPPAttack  & 0.945 & 1.000 & 0.907 & 0.024 & 0.035 & 0.987 & 0.907 & 0.012 \\
& Trojan     & 0.947 & 1.000 & 0.799 & 0.113 & 0.272 & 0.691 & 0.837 & 0.036 \\
& LF         & 0.944 & 0.997 & 0.075 & 0.944 & 0.689 & 0.211 & 0.748 & 0.111 \\
& Poison-Ink & 0.946 & 0.998 & 0.784 & 0.165 & 0.127 & 0.880 & 0.854 & 0.058 \\

\bottomrule
\end{tabular}
}
\end{table*}

\vspace{-4mm}

\section{Experiments}

\subsection{Experimental Setup}

\noindent \textbf{Dataset and Model Settings}
To demonstrate the effectiveness of \texttt{Lite-BD}, we evaluate our approach on four widely used datasets: CIFAR-10 \cite{krizhevsky2009learning}, Fashion-MNIST \cite{xiao2017fashionmnist}, GTSRB \cite{stallkamp2011german}. We employ two commonly used architectures, ResNet-18 \cite{he2016deep} and VGG-11 \cite{simonyan2014very}. Specifically, CIFAR-10 and Fashion-MNIST are classified using ResNet-18, while GTSRB is evaluated using VGG-11. We also extend our evaluation on CIFAR-10 using MobileNetV1 \cite{mobilenets2017}, DenseNet-121 \cite{huang2017densely}, and Vision Transformer (ViT-Base) \cite{dosovitskiy2020vit} models to demonstrate the generalizability of our method across diverse architectures. 

\noindent \textbf{Evaluation Metrics}
To assess the effectiveness of our method and enable fair baseline comparisons, we adopt three evaluation metrics: (1) \emph{Clean Accuracy (CA)}, which measures the classification accuracy on clean samples after passing through the defense framework; (2) \emph{Poisoned Accuracy (PA)}, which denotes the classification accuracy of poisoned samples after defense; and (3) \emph{Attack Success Rate (ASR)}, which represents the proportion of poisoned samples that successfully trigger the backdoor attack. An effective defense is characterized by higher CA, higher PA, and lower ASR.

\noindent \textbf{Attack Methods}
We evaluate our defense framework against ten state-of-the-art (SOTA) backdoor attacks, including BadNets \cite{gu2019badnets}, Blend \cite{chen2017targeted}, WaNet \cite{nguyen2021wanet}, SIG \cite{barni2019new}, Clean-Label (CL) \cite{turner2018clean}, BPPAttack \cite{wang2022bppattack}, Trojan \cite{liu2018trojaning}, LF \cite{zeng2021rethinking}, Poison-Ink \cite{zhang2022poisonink}, and LIRA \cite{doan2021lira}.  In Figure \ref{fig:purified_samples_visulization} poisoned samples from all the attacks with their corresponding triggers are shown.

\noindent \textbf{Baseline Defense Methods}
We compare our proposed defense framework with three SOTA black-box defense methods: (1) \emph{ShrinkPad} \cite{li2021backdoor}, which disrupts backdoor triggers through image shrinking and padding; (2) \emph{ZIP} \cite{Shi2023ZIP}, which applies linear transformations such as blurring to disrupt backdoor triggers and subsequently restores purified images using diffusion models; and (3) \emph{SampDetox} \cite{Yang2024SampDetox}, a recent black-box backdoor defense that leverages noise-denoising–based detoxification with diffusion models to purify poisoned images.

\vspace{-3mm}

\subsection{Main Results}

The performance of the \texttt{Lite-BD} and its comparison with state-of-the-art black-box backdoor defenses are presented in Table~\ref{tab:defense_comparison}. For CIFAR-10, \texttt{Lite-BD (RE)} achieves the highest average CA and PA, alongside the lowest average ASR. Specifically for BadNets, the CA is marginally lower ($\approx 2\%$) than the top-performing ZIP; however, both ZIP and SampDetox fail completely to mitigate the attack. In our attack design, the BadNets trigger size is $6 \times 6$, which the diffusion models in ZIP and SampDetox may reconstruct, leading to defense failure. For Poison-Ink, our method achieves the second-best CA and the third-lowest ASR. On Fashion-MNIST, the \texttt{Lite-BD (RE)} yields the lowest average ASR. By considering the results of both \texttt{Lite-BD (RE)} and \texttt{Lite-BD (SW)}, we achieve the best CA in two attacks, the best PA in three, and the lowest ASR in six scenarios. Furthermore, the gap between the best-reported metrics and our proposed method remains within $5\%$ across four attacks. For GTSRB, \texttt{Lite-BD (RE)} achieves the lowest average ASR, obtaining the best CA in four attacks and the lowest ASR in six. While ZIP and SampDetox achieve higher CA and PA in some GTSRB cases, it is critical to note that their diffusion models are trained on the evaluation datasets themselves. In contrast, our method utilizes pre-trained models trained on entirely disparate datasets. Despite this disparity, our method remains competitive, achieving comparable CA and PA in most cases. Beyond defensive robustness, a primary advantage of our approach is computational efficiency. As demonstrated in Table~\ref{tab:execution_time}, our method significantly reduces execution time per sample. On average for CIFAR-10, \texttt{Lite-BD (RE)} and \texttt{Lite-BD (SW)} are \textbf{123.4$\times$} and \textbf{102.8$\times$} faster than Sampdetox, and \textbf{34$\times$} and \textbf{28.33$\times$} faster than ZIP, respectively. Similar gains are observed for Fashion-MNIST and GTSRB. This analysis confirms that \texttt{Lite-BD} is efficient, dataset-independent, and achieves performance comparable to or exceeding state-of-the-art black-box defenses. The purified samples from backdoor triggers for all the attacks are illustrated in Figure \ref{fig:purified_samples_visulization}.

\begin{table}[htbp]
\centering
\small
\caption{Performance on More Models. This experiment is conducted using CIFAR-10.}
\label{tab:more_models}
\resizebox{\linewidth}{!}{%
\begin{tabular}{lcccccc}
\toprule
\textbf{Model} & \textbf{Attack} & \textbf{Baseline CA} & \textbf{Baseline ASR} & \textbf{Defense PA} & \textbf{Defense ASR} \\
\midrule
\multirow{5}{*}{DenseNet-121}
& BadNet & 0.891 & 1.000 & 0.895 & 0.003 \\
& Blend  & 0.893 & 0.997 & 0.862 & 0.012 \\
& WaNet  & 0.891 & 0.991 & 0.882 & 0.025 \\
& SIG    & 0.825 & 0.996 & 0.598 & 0.086 \\
& LF     & 0.885 & 0.987 & 0.706 & 0.078 \\
\midrule
\multirow{5}{*}{MobileNetV1}
& BadNet & 0.694 & 1.000 & 0.777 & 0.006 \\
& Blend  & 0.770 & 0.993 & 0.733 & 0.008 \\
& WaNet  & 0.794 & 0.992 & 0.781 & 0.066 \\
& SIG    & 0.729 & 0.999 & 0.262 & 0.004 \\
& LF     & 0.753 & 0.981 & 0.579 & 0.128 \\
\midrule
\multirow{5}{*}{ViT-Base}
& BadNet & 0.736 & 1.000 & 0.782 & 0.175 \\
& Blend  & 0.735 & 0.988 & 0.680 & 0.014 \\
& WaNet  & 0.686 & 0.979 & 0.655 & 0.052 \\
& SIG    & 0.664 & 0.621 & 0.265 & 0.203 \\
& LF     & 0.671 & 0.972 & 0.571 & 0.005 \\
\bottomrule
\end{tabular}
}
\end{table}

\vspace{-8pt}

\subsection{Impact of Different Stages of the Defense Framework}

Table~\ref{tab:comprehensive_ablation} illustrates the individual contributions of different stages of the \texttt{Lite-BD (RE)} across multiple backdoor attacks. From the results, we observe that most attacks can be effectively mitigated by \textbf{Stage~1}, except for \textbf{SIG} and \textbf{LF}, which require the involvement of \textbf{Stage~2}. When the trigger contains high-frequency components distributed across the spatial dimensions, as in SIG, or exhibits smooth transitions over the entire image, as in LF, band-by-band frequency filtering becomes more effective. Due to space limitations, we only present results on CIFAR-10; however, similar trends are observed on GTSRB and Fashion-MNIST.

\vspace{-3mm}

\subsection{Impact of Different Models on the Defense Framework}

Table~\ref{tab:more_models} reports results on additional architectures, including DenseNet-121, MobileNetV1, and ViT-Base, under five backdoor attacks on CIFAR-10. The results demonstrate consistently strong defense performance across nearly all models and attacks, indicating that our method generalizes well across architectures. These findings confirm that our defense framework is \textbf{model-agnostic}.

\begin{figure}
    \centering
    \includegraphics[width=\linewidth]{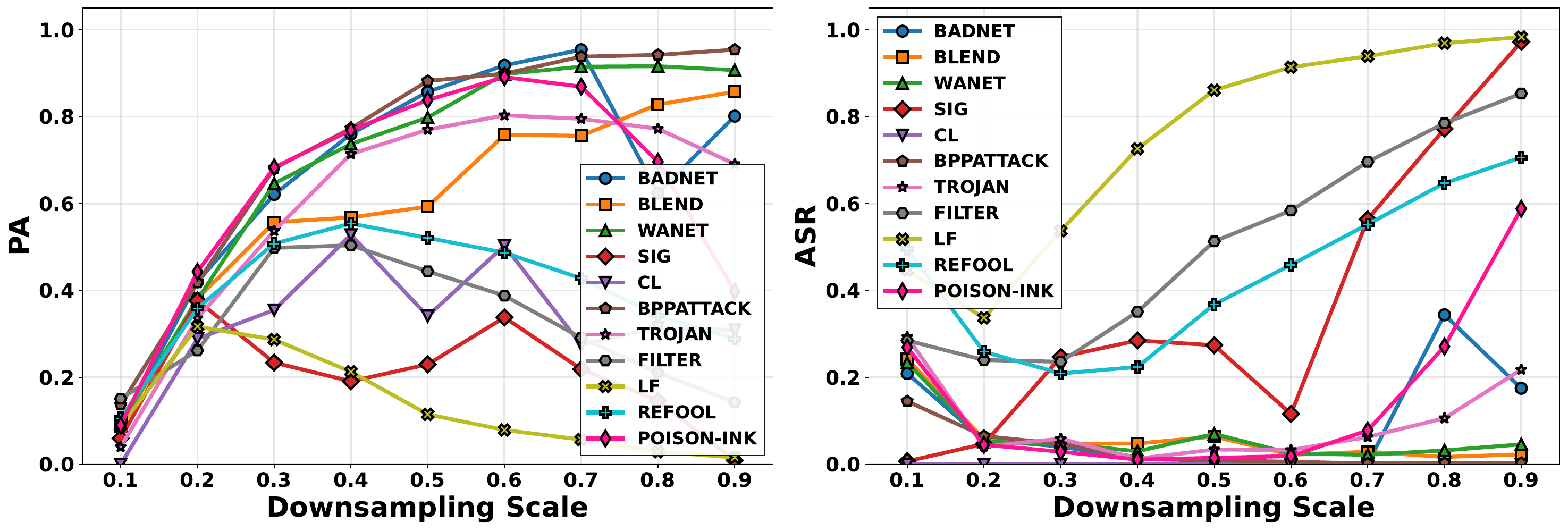}
    \caption{Defended PA and ASR for different down-sampling scale of Stage~1. This experiment was done on CIFAR-10.}
    \label{fig:downsample}
\end{figure}

\vspace{-3mm}

\subsection{Impact of the Downscaling Ratio on Defense Performance}

Figure~\ref{fig:downsample} illustrates the impact of the downscaling ratio $s$ on prediction accuracy (PA) and attack success rate (ASR). As expected, smaller values of $s$ lead to lower PA and lower ASR, while larger values of $s$ yield higher PA and higher ASR. Therefore, the defender should select an intermediate value of $s$ that balances maintaining high PA while effectively suppressing ASR. Due to space limitations, we only present results on CIFAR-10; however, similar trends are observed on GTSRB and Fashion-MNIST.

\vspace{-3mm}

\section{Conclusion}

\vspace{-3mm}

In this paper, we propose a novel black-box backdoor defense method called \texttt{Lite-BD} that utilizes down-upscaling transformation with a pre-trained super-resolution network and a query-based band-by-band frequency filtering technique to disrupt backdoor triggers while maintaining the benign features of poisoned samples. Our method is lightweight, zero-shot, and considers both spatial and frequency transformations that are absent in state-of-the-art black-box defenses.

\vspace{-3mm}

\bibliographystyle{plain}
\bibliography{custom}

\end{document}